\documentclass[aps,preprint,nofootinbib]{revtex4-1}
\usepackage{amsmath,bm}
\usepackage{graphicx}
\usepackage{enumerate}
\usepackage{hhline}
\usepackage{graphicx,color}
\usepackage{amsmath}
\usepackage{amsfonts}
\usepackage{amssymb}
\usepackage{amsfonts}
\usepackage{graphicx}
\usepackage{subfigure}
\usepackage{color}
\usepackage{setspace}
\usepackage[toc,page]{appendix}

\newcommand{\B}[2]{B^{#1}{}_{#2}}   			

\newcommand{\torsion}[3]{T^{#1}{}_{#2#3}}

\newcommand{\s}[3]{S^{#1}{}_{#2#3}}

\begin{document}

\title{Parametrized Post-Newtonian Limit of  Teleparallel Dark Energy Model}

\author{Jung-Tsung Li}

\email{s100022519@m100.nthu.edu.tw}

\author{Yi-Peng Wu}

\email{s9822508@m98.nthu.edu.tw}

\author{Chao-Qiang Geng}

\email{geng@phys.nthu.edu.tw}

\affiliation{Department of Physics, National Tsing Hua University, Hsinchu 300,
Taiwan\\
Physics Division, National Center for Theoretical Sciences, Hsinchu
300, Taiwan}

\date{\today}


\begin{abstract}
We study the post-Newtonian limit in the \textit{teleparallel equivalent of General Relativity}
 with a scalar field which non-minimally couples to gravity. The metric perturbation is obtained from the vierbein
 field expansion with respect to the Minkowski background. Due to the structure of the teleparallel gravity Lagrangian,
 the potential of the scalar field shows no effect to the parametrized post-Newtonian parameters, and 
compatible results with Solar System observations are found. 
 \end{abstract}
\maketitle


\section{Introduction}
Teleparallelism, originally introduced by Einstein~\cite{A. Einstein} for
the unification of gravity and electromagnetism, is now regarded as 
an attractive alternative theory for
  interpreting the classical gravitational effect.
While the notion of parallelism is missing for vectors in general relativity,  
it is easy to consider two vectors separating at a finite distance to be ``parallel''
in teleparallel gravity.
By using the curvatureless Weitzenb\"{o}ck connection instead of 
the torsionless Levi-Civita connection, the Einstein-Hilbert Lagrangian of gravity can be formulated, 
%
which is known as 
the \textit{Teleparallel Equivalent of General Relativity }(TEGR)~\cite{F.W. Hehl, K. Hayashi, E.E. Flanagan, J. Garecki}.

One can formulate the teleparallel dark energy model~\cite{Geng:2011aj, Geng:2013} by adding a canonical scalar field 
with a non-minimal coupling term in TEGR. When the non-minimal coupling parameter is zero, it is  
identical to the ordinary quintessence-like theory. However, when the coupling is turned on, the model has 
different behaviors in comparison with GR. 
In particular, it has been found that the cosmic acceleration is driven in this class of models even when the potential
of the scalar field is neglected~\cite{Gu:2013}.
Given that a very light field in the scalar-tensor theory of GR can significantly
modify
the results in the weak field regime~\cite{Olmo}, 
 it is thus essential to check if the non-minimally interacted theory in teleparallel gravity can 
fit the Solar System 
observations~\cite{Reasenberg:1979ey, Fomalont:2009zg, Bertotti:2003rm, Williams:2004qba}.

The framework for analyzing the weak-field limit~\cite{Wagoner:1970vr,Wagoner:1976am} is the parametrized post-Newtonian (PPN) 
formalism. The original version was studied by Eddington~\cite{eddington}, Robertson~\cite{robertson}, 
and Schiff~\cite{schiff}, who introduced the metic 
	\begin{equation}
	ds^2= - \left( 1-2{GM\over rc^2} + 2\beta \left({GM \over rc^2}\right)^2  \right) dt^2 
	+ \left( 1+2\gamma {GM\over rc^2} \right) dr^2+r^2d\Omega^2,
	\end{equation}
where $M$ is the mass of the Sun, $r$ is the radial coordinate,  and $\beta$ and $\gamma$
measures the amounts of the nonlinearity $(M/rc^2)^2$ and  
curvature,
respectively. 
In GR, $\beta=\gamma=1$, while other theories may predict small deviations from unity 
due to the influences of other fields. 
Experimentally, we have that
$\gamma-1=\left( 2.1\pm2.3 \right)\times 10^{-5}$ from 
the Cassini experiment~\cite{Bertotti:2003rm} and
$\beta-1=\left(1.2\pm1.1\right)\times 10^{-4}$
from the lunar laser ranging tests~\cite{Williams:2004qba}.

We  aim to calculate the weak-field expansion in TEGR with a scalar field 
which non-minimally couples to gravity. By noting that the dynamic variable is the vierbein field, 
we  relate
 the tetrad field to the metric in order to obtain the 
PPN values. 
The post-Newtonian approach of the matter source in 
teleparallel gravity is similar to the GR case~\cite{Chandrasekhar:1965, will}. 
Conventionally, the velocity of the testing object $v=\sqrt{GM/r}$ characterizes the smallness of the system 
in powers of $v/c$.
With the accuracy of the metric to $O(v^2/c^2)\equiv O(2)$, which is the Newtonian limit, 
the energy momentum tensor can be determined to a sufficient order 
to obtain the post-Newtonian terms $g_{00}$ to $O(4)$ and $g_{0i}$ to $O(3)$.

In this work, 
we first study the weak-field regime of teleparallel gravity by
  expanding the vierbein field around the Minkowski background. 
After solving the equation of motion in a slow-moving system, we find that the gravitational effect
can be attributed to a potential from torsion caused by the source matter which shares the same form as the Newtonian potential.
We then solve the field equation of TEGR, including a non-minimally coupled scalar field, up to the post-Newtonian limit
in terms of the gravitational force potential. 
The resulting metric tensor obtained from the perturbed vierbein field gives PPN
parameters, 
which can be compared with the observational data 
and used to constrain the model. 

The paper is organized as follows. In Sec. II, we briefly introduce the math tools of teleparallel gravity and  
add a scalar field with the non-minimal coupling term in the action. The required knowledge of 
the PPN formalism is given in Sec. III. In Sec. IV, we perturb the field equation and obtain the post-Newtonian limit in tetrad fields. 
The summary is presented in Sec. V.
\section{Teleparallel Gravity}


For a spacetime $(M,g)$, a vector in the tangent space $T_p M \cong
\mathbb{R}^{1,3}$ can be spanned by the coordinate vector fields 
$\{\partial_\mu \}$. The covector  basis $\{ dx^\mu \}$ describes the cotangent space 
$T^*_p M\cong \mathbb{R}^*{}^{1,3}$ dual to $T_pM$. Meanwhile,
we may also choose an orthonormal basis (frame) $\{ e_a
|a=0,\ldots,3\}$ to span $T_pM$,
in which 
$g\left( e_a, e_b \right)= \eta_{ab}$. 
This orthonormal basis $\{ e_a \}$ is called  \textbf{tetrad} or
\textbf{vierbein}, standing for four-legged. 
We denote the dual basis of $\{e_a\}$ by $\{ e^a \in T_p^*M\}$,
satisfying $e^a(e_b) = \delta_b^a$. Both frames can be
written in terms of the local coordinate basis by
\begin{equation}
\label{eq: expansion}
e_a=e_a{}^\mu \partial_\mu \quad \text{and} \quad e^a=e^a{}_\mu dx^\mu,
\end{equation}
such that $\delta^a_b= e_a{}^\mu e^b{}_\mu$,
and conversely
	\begin{equation}
	\label{eq: vierbein orthogonal}
		e^a{}_\mu e_a{}^\nu=\delta_\mu^\nu.
	\end{equation}
In the following discussions, we shall use Greek indices
$\mu,\nu,\gamma,\ldots = 0,\ldots,3$  and Latin
indices $a,b,c,\ldots = 0,\ldots,3$ for the coordinate  and   orthonormal bases, respectively.
The metric is related to a vierbein field 
through
$g_{\mu\nu}e_a{}^\mu e_b{}^\nu= \eta_{ab}$,
or inversely
\begin{equation}
\label{eq: orthogonality}
g_{\mu\nu}=\eta_{ab}e^a{}_\mu e^b{}_\nu.
\end{equation}

In teleparallel gravity, one uses the Weitzenb\"{o}ck connection
~\cite{weitzenbock}, which allows the
path-independent parallel transport 
$\overset{w}{\nabla}_{e_a}e_b{}^\nu = e_a{}^\mu\overset{w}{\nabla}_\mu e_b{}^\nu =0$. This is the notion of \textit{absolute
parallelism} or \textit{teleparallelism}, and the
connenction reads
\begin{equation}
\overset{w}{\Gamma}{}^{\nu}{}_{\mu\lambda}=  - e^b{}_\lambda\partial_\mu e_b{}^\nu = e_b{}^\nu\partial_\mu e^b{}_\lambda.
\end{equation}
Note that the two lower indices $\mu$ and $\nu$ 
are in general not symmetric, and thus the torsion tensor is nonzero in the teleparallel spacetime.  
The components of the Weitzenb\"{o}ck torsion tensor in the coordinate basis are 
\begin{equation}
T^\lambda{}_{\mu\nu}
= \overset{w}{\Gamma}{}^\lambda{}_{\nu\mu} -\overset{w}{\Gamma}{}^\lambda{}_{\mu\nu} 
= e_a{}^\lambda \left(  \partial_\nu e^a{}_\mu - \partial_\mu e^a{}_\nu  \right)  .
\end{equation}

In TEGR, the gravitational action is composed of the square of the Weiteznb\"{o}ck torsion tensor 
instead of the Ricci scalar. Teleparallel gravity coincides with the Einstein's theory built from the metric when 
the teleparallel Lagrangian density is 
\begin{equation}
 T\equiv
\frac{1}{4}\ T^{\rho}{}_{\mu\nu}T_{\rho}{}^{\mu\nu}+\frac{1}{2}\ T^{\rho}{}_{\mu\nu}
T^{\nu\mu}{}_{\rho}-T^{\rho}{}_{\mu\rho}T^{\nu\mu}{}_{\nu}.
\end{equation}
The dynamics on teleparallel gravity is described by
\begin{equation}
\label{eq: TEGR action}
\mathcal{S}= 
\int d^4 x e\bigg[{T\over 2\kappa^2}+\mathcal{L_M}\bigg],
\end{equation}
where $e=det(e^a{}_\mu)= \sqrt{-g}$ and $ \kappa^2 = 8\pi G/c^4$. 
By variation with respect to the dynamical variable $e_a{}^{\mu}$, the equation of motion is given by
\begin{equation}
\label{eq: TEGR fieldeq}
e^{-1}\partial_\mu\left( ee_a^\lambda S_\lambda{}^{\mu\nu} \right) + e_a{}^\lambda S_\rho{}^{\nu\mu} 
T^\rho{}_{\mu\lambda} + {1\over 4} e_a{}^\nu S^\rho{}_{\mu\nu} T_\rho{}^{\mu\nu} 
= {4\pi G\over c^4} e_a{}^\mu \mathcal{T}_\mu{} ^\nu,
\end{equation}
where $\mathcal{T}_\mu{}^\nu$ is the usual matter energy momentum tensor, and  $S^\rho{}_{\mu\nu}$ 
is defined according to 
\begin{equation}
S^\rho{}_{\mu\nu} 
= {1\over4} \left( T^\rho{}_{\mu\nu}- T_{\mu\nu}{}^\rho + T_{\nu\mu}{}^\rho   \right) 
+ {1\over2} \delta^\rho_\mu T_{\sigma\nu}{}^\sigma - {1\over2} \delta^\rho_\nu T_{\sigma\mu}{}^\sigma ,
\end{equation}
such that the torsion scalar is given by $T=S^\rho{}_{\mu\nu} T_\rho{}^{\mu\nu}$.
It is noteworthy that Eq.~(\ref{eq: TEGR fieldeq}) is a system of 16 equations since $e^a_\mu$ 
in general is an asymmetric tensor field. Thus, we can construct 
 the vierbein $e^a_\mu$ at each point through the matter distribution and use the orthogonality relation in 
 Eq.~(\ref{eq: orthogonality}) to obtain the metric. By the transition to coordinate, nevertheless, one finds that
 Eq.~(\ref{eq: TEGR fieldeq}) is nothing but the Einstein field equation of GR. 
 This implies that only 10 degrees of freedom in the vierbein are relevant in the TEGR theory.
 %

For the teleparallel dark energy model in which 
 a canonical scalar field non-minimally couples to TEGR gravity,
 the Lagrangian reads~\cite{Geng:2011aj}
	\begin{equation}
	\label{eq: nonminimal action}
		\mathcal{S}= 
		\int d^4 x e\bigg[{T\over 2\kappa^2}+ {1\over2}\left( \partial_\mu \phi  \partial^\mu \phi 
		 + \xi T 			\phi^2  \right) - V(\phi)  +\mathcal{L_M}\bigg],
	\end{equation}
with $V(\phi)$ the scalar field potential and $\xi$ the non-minimal coupling parameter. 
In the case without the minimal coupling, $i.e.$ $\xi=0$, it is an ordinary quintessence type. 
When $\xi$ is not zero, the non-minimal coupling term $ \xi T \phi^2 $ results in
a distinct  dark energy model~\cite{Geng:2011aj}. Variating with respect to the vierbein and scalar field, 
we have the following equations of motion:
	\begin{eqnarray}
	\label{eq: nonminimal fieldeq}
	& &  \left( {c^4\over 4\pi G} + 2\xi \phi^2 \right) \bigg[ e^{-1}\partial_\mu\left( ee_a{}^\lambda S_\lambda{}^{\mu\nu} \right) 
	+ 	e_a{}^\lambda S_\rho{}^{\nu\mu} T^\rho{}_{\mu\lambda} 
	+ {1\over 4} e_a{}^\nu S^\rho{}_{\mu\nu} T_\rho	{}^{\mu\nu}  \bigg] \nonumber\\
	& &  -e_a{}^\nu \bigg[ {1\over2}  \partial_\mu \phi  \partial^\mu \phi 
	- V(\phi)  \bigg] + e_a{}^\mu\partial^\nu	\phi\partial_\mu\phi
	+ 4\xi e_a{}^\rho S_\rho{}^{\mu\nu} \phi    \left(  \partial_\mu \phi  \right) = e_a{}^\rho \mathcal{T}_\rho{}^\nu,
	\end{eqnarray}
and 
	\begin{equation}
	\label{eq: scalar field fieldeq}
		\Box \phi - \xi T \phi + {dV\over d\phi} = 0,
	\end{equation}
with $\Box \equiv g^{\mu\nu} \nabla_\mu \nabla_\nu$.
Here, once again, Eq.~(\ref{eq: nonminimal fieldeq}) is composed by 16 equations and the transition to coordinate 
leads to 6 constraint equations, given by
	\begin{equation}
	\label{eq: TDE constaints}
		4 \xi \phi \left(S_\lambda{}^{\mu\nu}-S^{\nu\mu}{}_{\lambda}\right)\partial_\mu\phi = 0,
	\end{equation}
which vanish in the TEGR limit $\xi=0$. These constraints significantly modify the evolution history of the density perturbations
on all scales of the universe~\cite{Geng:2013}.


\section{the Post-Newtonian Approach}
\label{sec: PPN Approach}
This section is a generalization of the post-Newtonian formalism from Ref.~\cite{will} 
in order to have the coherent form as other theories of gravity. In the perturbation theory, 
the order of the smallness of each function form is the key to determine the magnitudes of terms in field equations. 
Typically, the Newtonian gravitational potential $U/c^2$ is nowhere larger than $10^{-5}$. 
The velocity of the fluid element is related to $U$ by the viral theorem $\mathnormal{v^2} \sim U$. 
The fluid pressure inside the Sun is generally smaller than the 
energy density $\rho U$, 
which can be expressed as $p/\rho \lesssim U$ ($p/\rho\sim 10^{-5}$ in the Sun). 
Besides, one must consider the internal energy $\rho\Pi$ in the Solar System,
in which the specific energy density $\Pi$ is related to $U$ by $\Pi \lesssim U$ ($\Pi/c^2$ is $\sim 10^{-5}$ in the Sun). 
The order of the smallness is given by
\begin{equation}
{U\over c^2} \sim {\mathnormal{v^2} \over c^2} \sim {p\over \rho c^2} \sim {\Pi \over c^2} \sim O(2).
\end{equation}
 In addition, the time derivative is related to the motion of its constituents through $\partial/\partial t \sim \mathsf{v \cdot \nabla}$, 
 so that the order of the smallness relative to the spatial derivative is 
\begin{equation}
\frac{\lvert \partial/\partial t\rvert}{\lvert \partial/\partial x\rvert}  \sim O(1).
\end{equation}
The perfect fluid has the energy momentum of the form
\begin{equation}
\label{eq: em tensor}
\mathcal{T}_{\mu\nu}= \left( \rho c^2 + \rho \Pi +p \right) u_\mu u_\nu+p g_{\mu\nu}, 
\end{equation}
where $u^\mu$ is the covariant four-velocity defined as $u^\mu= dx^\mu/ds$.
To the relevant order, the components of the energy momentum are
	\begin{subequations}
	\label{eq: em expansion}
	\begin{align}
		& \mathcal{T}_{00}= \rho c^2 \left( 1+\frac{1}{c^2}  \left( v^2+2U+\Pi \right) \right) + \ldots\,,\\
		& \mathcal{T}_{0i} = -\rho c v_i + \ldots\,,\\
		& \mathcal{T}_{ij}  = \rho v_i v_j +\delta_{ij}p + \ldots\,.
	\end{align}
	\end{subequations}
Here, we have adopted the approximations 
$u_0=- 1-(v^2/2+U)/c^2 \,+ \,O(4)$ and $u_i=v_i/c\,+\,O(3)$ 
to obtain the results in Eq.~(\ref{eq: em expansion})~\cite{Chandrasekhar:1965}.  

%
For the gravitational sector, we expand the vierbein field around the flat background as
    \begin{equation}
		e^a{}_\mu=\delta^a{}_\mu + {}^{(2)}B^a{}_\mu +{}^{(3)}B^a{}_\mu+{}^{(4)}B^a{}_\mu+ \ldots\,,
	\end{equation}
where ${}^{(2)}B^a{}_\mu$ denotes $B^a{}_\mu$ to $O(2)$ and so on.
Here, without loss of generality, we align the background orthonormal basis with the coordinate $e^a=\delta^a{}_\mu dx^\mu$,
but the breakdown of such alignment is allowed at the perturbation level. 
This decomposition gives the usual metric as $g_{\mu\nu}=\eta_{\mu\nu}+h_{\mu\nu}$, and to our purpose
it suffices to consider up to $O(4)$,  where
    \begin{eqnarray}
    \label{eq: linear weak metric}
{}^{(2+3)}h_{\mu\nu}&=&\eta_{ab}\left(\delta^a{}_\mu{}^{(2+3)}B^b{}_\nu+{}^{(2+3)}B^a{}_\mu\delta^b{}_\nu\right),\\
{}^{(4)}h_{\mu\nu}&=&\eta_{ab}\left(\delta^a{}_\mu{}^{(4)}B^b{}_\nu+{}^{(4)}B^a{}_\mu\delta^b{}_\nu\right)+\eta_{ab}{}^{(2)}B^a{}_\mu{}^{(2)}B^b{}_\nu.
    \label{eq: PPN metric}    
    \end{eqnarray}
With in mind that ${}^{(4)}h_{00}$ is the only relevant term in the PPN approach at $O(4)$, one observes that
${}^{(2)}B^a{}_0$ may contribute to the perturbed metric. 
By denoting $B^\rho{}_\mu=\delta^\rho_a B^a{}_\mu$,  we now raise and lower the spacetime indices by the Minkowski metric $\eta_{\mu\nu}$:
	\begin{equation}
 		B_{\nu\mu}=\eta_{\nu\rho} B^\rho{}_\mu.
	\end{equation}
In the Newtonian limit, the $O(2)$ relation of Eq.~(\ref{eq: linear weak metric}) implies
${}^{(2)}h_{00}=2{}^{(2)}B_{00}$, ${}^{(2)}h_{ij}={}^{(2)}B_{ij}+{}^{(2)}B_{ji}$ for $i=j$, as well as two constraints
    \begin{equation}
    {}^{(2)}h_{0i}={}^{(2)}B_{0i}+{}^{(2)}B_{i0}=0,\;\;\;\;\;\;{}^{(2)}h_{ij}={}^{(2)}B_{ij}+{}^{(2)}B_{ji}=0,
    \end{equation}
for $i\neq j$, which illustrate the 6 decoupled degrees of freedom in the perturbed vierbein field.

If only standard results are expected in the PPN limit, as shall be the case of TEGR, the non-vanished perturbations of $B_{\mu\nu}$ 
exist as follows:
	\begin{eqnarray}
	\label{eq: tetradperturbation}
		 B_{00} &=& {}^{(2)}B_{00} + {}^{(4)}B_{00} + O(6) , \nonumber\\
		 B_{0i} &=& {}^{(3)}B_{0i} + O(5) , \nonumber\\
		 B_{i0} &= & {}^{(3)}B_{i0} + O(5) , \nonumber\\
		 B_{ij} &=& {}^{(2)}B_{ij} + O(4) .
	\end{eqnarray}    
As will be demonstrated in the next section, we find that Eq.~(\ref{eq: tetradperturbation}) shows the solution
in the rest frame of the Sun with a diagonal vierbein in the Newtonian limit where ${}^{(2)}B_{0i}={}^{(2)}B_{ij}=0$ for $i\neq j$.
Moreover, we shall also adopt the gauge conditions~\cite{Smalley:1980em,Hobson}:
    \begin{eqnarray}
    \label{eq: ppn gauge}
       B_i{}^\mu{}_{,\mu} + B^\mu{}_i{}_{,\mu} - B_\mu{}^\mu{}_{,i}&=&0 ,\nonumber\\
       B_0{}^\mu{}_{,\mu}+ B^\mu{}_0{}_{,\mu} - B_\mu{}^\mu{}_{,0}&=&- B_{00,0},
    \end{eqnarray}
    in which we directly derive the usual PPN gauge formulas~\cite{Will:1971zzb}, 
    \begin{eqnarray}
      h_i{}^\mu{}_{,\mu}-{1\over 2} h_\mu{}^\mu{}_{,i}&=&0 ,\nonumber\\
      h_0{}^\mu{}_{,\mu}-{1\over 2} h_\mu{}^\mu{}_{,0}&=&-{1\over 2} h_{00,0},
   \end{eqnarray}
by using Eq.~(\ref{eq: tetradperturbation}) to our gauge choice in Eq.~(\ref{eq: ppn gauge}).
%


\section{the post-Newtonian limit of teleparallel gravity}

In teleparallel gravity, since gravitation is attributed to torsion,  
there are no geodesic equations, but only force equations similar to the Lorentz force equations 
of electrodynamics~\cite{de Andrade:1997qt, Pereira}. 
The equation of motion can be written in the purely spacetime form
	\begin{equation}
	{du^\rho \over ds}+ \overset{w}{\Gamma}{}^{\rho}{}_{\mu\nu}u^\mu u^\nu
	 = T_\mu{}^\rho{}_\nu u^\mu u^	\nu\,.
	\end{equation} 
In the Newtonian limit, the Weitzenb\"{o}ck connection reads 
${}^{(2)}\overset{w}{\Gamma}{}^{\rho}{}_{\mu\nu}=\partial_\nu {}^{(2)}B^\rho{}_\mu$.
With the vanishingly small velocities $u^i$ and weak gravitational field $\lvert B^a{}_\mu \rvert \ll 1$, the force equation reduces to 
\begin{equation}
{d^2x^i \over dt^2} = - \partial^i U   \simeq c^2 \partial^i {}^{(2)}B_{00},
\end{equation}
where $U$ is the gravitational potential determined in terms of $\rho$ by the Poisson's equation, 
and we consider $U=-G\int d^3 x'\rho(\mathbf{x'},t)/\lvert \mathbf{x-x'} \rvert$.
As a result, we have ${}^{(2)}B_{00} = -U/c^2$, which gives the correct Newtonian metric ${}^{(2)}h_{00} = -2U/c^2$.

\subsection{the equivalence of general relativity}

So far we have obtained the standard result up to $O(2)$ without specifying any gravitational Lagrangian in teleparallel gravity.
It is now straightforward to check whether our formalism can lead to PPN parameters identical to GR in the equivalent description.
To proceed the calculation, we rewrite the TEGR field equation in Eq.~(\ref{eq: TEGR fieldeq}) in the following form: 
\begin{equation}
\label{eq: fieldeq2}
e^a{}_\mu e^{-1}\partial_\rho(eS_a{}^{\rho\nu})-\frac{1}{2}\delta_\mu^\nu e^a{}_\sigma e^{-1}
\partial_\rho \left( eS_a{}^{\rho\sigma}  \right)+S_\rho{}^{\nu\sigma}
T^\rho{}_{\sigma\mu}+\frac{1}{4}\delta_\mu^\nu T 
= {4\pi G\over c^4} \left(  \mathcal{T}_\mu{}^\nu-\frac{1}{2}\delta_\mu^\nu \mathcal{T}  \right),
\end{equation}
where $\mathcal{T}$ is the trace of the energy momentum tensor
(see Appendix A.), and we linearize the above equation by Eqs.~(\ref{eq: em expansion}) and (\ref{eq: tetradperturbation}) with ${}^{(2)}B_{00} = -U/c^2$.
Note that the torsion tensor $T^\rho{}_{\sigma\mu}$ is at least an $O(2)$ quantity, and consequently,
the third and fourth terms on the left-hand side of Eq.~(\ref{eq: fieldeq2}) must have the order of magnitude to be at least 
 $O(4)$, which only involve in the computation of ${}^{(4)}B_{00}$.
 
Let us begin by perturbing the $(i, j)$ component of Eq.~(\ref{eq: fieldeq2}), where only $O(2)$ terms are kept.
The linearized field equation shows simply
	\begin{equation}
	\label{eq: fieldeqij}
		\partial^\rho S_{i\rho}{}^j-{1\over2}\delta_i^j\partial^\rho S^\sigma{}_{\rho\sigma} 
		= \frac{4\pi G}{c^4}  \left( \mathcal{T}_i{}^j-{1\over2}  \delta_i^j \mathcal{T}  \right).
	\end{equation}
Taking the trace of Eq.~(\ref{eq: fieldeqij}) and picking up the PPN gauge conditions 
in Eq.~(\ref{eq: ppn gauge}), we can reduce the above equation into the Poisson's equation:
	\begin{equation}
	\begin{aligned}
		 -\nabla^2 B^i{}_i = {4 \pi G\over c^4}\left( 3\rho c^2  \right)\,,
	\end{aligned}
	\end{equation}
where $\nabla^2\equiv \delta_{ij}\partial_i\partial_j$.
The integration of this Poisson's equation is nothing but $B^i{}_i=-3U/c^2$,
which implies
	\begin{equation}
		\label{eq: Bij}
		{}^{(2)}B^1{}_1={}^{(2)}B^2{}_2={}^{(2)}B^3{}_3=-{U\over c^2}.
	\end{equation}
As a result, one immediately observes the standard metric perturbation
\begin{equation}
\label{eq: h_ij}
{}^{(2)}h_{ij}={}^{(2)}B_{ij}+{}^{(2)}B_{ji}=-  {2U\over c^2} \delta_{ij}.
\end{equation}

Next, we  move to the $O(3)$ discussion where the $(0,i)$ component of the perturbed
Eq.~(\ref{eq: fieldeq2}) is taken into account. Given that $T^i{}_0\simeq \rho c v^i$ is of $O(3)$,
we have
	\begin{equation}
	\label{eq: fieldeq0i}
		\partial_\rho S_0{}^{\rho i}={4\pi G\over c^3}\rho v^i,
	\end{equation}
and the expansion of the left-hand side of Eq.~(\ref{eq: fieldeq0i}) is
	\begin{eqnarray}
		\partial_\rho S_0{}^{\rho i}  
		&=& -{1\over2}B^i{}_0 + {1\over2}\partial_0\partial_\rho B^{\rho i} 
		+ {1\over2}\partial^i \partial^\rho B_		{\rho 0} - {1\over2} \partial_0\partial^i B^{\sigma}{}_\sigma\nonumber\\
		&=& -{1\over2}B^i{}_0 + {1\over4c^3} \frac{\partial^2 U}{\partial t \partial x^i}\ .
	\end{eqnarray}
It is convenient to define new potentials 
$V_i$ and $W_i$~\cite{Chandrasekhar:1965}:
	\begin{subequations}
	\begin{align}
		& V_i\equiv G\int \frac{ \rho(\mathbf{x'},t)v'_i }{\lvert \mathbf{x-x'} \rvert} d^3x' ,\\
		& W_i \equiv G\int \frac{ \rho(\mathbf{x'},t)\mathbf{v' \cdot (x-x')} (x-x')_i }{\lvert \mathbf{x-x'} \rvert^3} 		d^3x' \,,
	\end{align}
	\end{subequations}
so that we obtain 
	\begin{equation}
	\label{eq: B_0i}
		{}^{(3)}B_{0i}=
	 {1\over c^3} \left( -{7\over 4}V_i - {1\over 4}W_i \right)\,,
	\end{equation}
while the similar study for the $(i,0)$ component shows
${}^{(3)}B^0{}_i=-{}^{(3)}B^i{}_0$ and ${}^{(3)}B_{0i}={}^{(3)}B_{i0}$,
corresponding to the perturbed metric
\begin{equation}
\label{eq: h_0i}
{}^{(3)}h_{0i}={}^{(3)}B_{0i}+{}^{(3)}B_{i0}={1\over c^3} \left( -{7\over 2}V_i - {1\over 2}W_i \right).
\end{equation}

Finally, we deal with the $(0,0)$ component of the perturbed equation motion up to $O(4)$.
Given that the relevant source at this order reads
\begin{equation}
\mathcal{T}_0{}^{0} -{1\over2} \delta^0_0 \mathcal{T}
= \frac{1}{2}\left( \rho c^2+ \rho\Pi + 2\rho v^2 +3p \right),
\end{equation}
it is useful to define the new potentials $\Phi_i\ (i=1,2,\cdots 4)$, where
	\begin{subequations}
	\begin{align}
	\label{potential Phi}
		\nabla^2\Phi_1= 4\pi G\rho v^2, \\
		\nabla^2\Phi_2= 4\pi G\rho U,  \\
		\nabla^2\Phi_3= 4\pi G\rho \Pi,  \\
		\nabla^2\Phi_4= 4\pi G p.
		\end{align}
	\end{subequations}	
As the detailed calculation can be found in 
 Appendix B, 
we only show  the solution of the perturbed vierbein here, that is
	\begin{equation}
	\label{eq: b_00}
	{}^{(2+4)}B_{00} = -{U\over c^2} - {U^2\over2 c^4} 
	-\frac{1}{c^4}\left( 2\Phi_1- 2\Phi_2 + \Phi_3 + 3\Phi_4 \right) . 
	\end{equation}	
The corresponding metric perturbation $h_{00}$ is given by Eqs.~(\ref{eq: linear weak metric}) and (\ref{eq: PPN metric}) as
	\begin{equation}
	\label{eq: h_00}
		{}^{(2+4)}h_{00} = -  {2U\over c^2} - {2U^2\over c^4} - 
		\frac{1}{c^4}\left(4\Phi_1- 4\Phi_2 + 2\Phi_3 +6\Phi_4\right) .
	\end{equation}
As seen from the results in Eqs.~(\ref{eq: h_ij}), (\ref{eq: h_0i}) and (\ref{eq: h_00}), we find that the vierbein formalism Eq.~(\ref{eq: tetradperturbation}) gives no deviation from the PPN limit of GR \cite{will}.

\subsection{the teleparallel dark energy model}

In the 
teleparallel dark energy model in Eq.~(\ref{eq: nonminimal action}), there is one additional
degree of freedom coming from the scalar field $\phi$. To study the PPN 
parameters, we  expand the scalar field 
as follows~\cite{Wagoner:1970vr}: 
	\begin{equation}
		\phi=\phi_0+\delta\phi^{(2)}+\delta\phi^{(4)} ,
	\end{equation}
where $\phi_0$ is a homogeneous background, 
which changes on the cosmological timescale, that is 
$\dot{\phi}_0\sim H_0/\kappa$ 
($H_0$ is the present Hubble constant).
The field equation in Eq.~(\ref{eq: scalar field fieldeq}) leads to
	\begin{equation}
	\label{eq: scalar field eq}
	\left( \eta^{\mu\nu}-h^{\mu\nu} \right) \left( \partial_\mu \partial_\nu \phi - \Gamma^\lambda{}_{\mu\nu}
			\partial_\lambda \phi  \right) - \xi T \phi_0 
			+ V''(\phi_0) \delta \phi + {1\over2}V'''(\phi_0) {\delta\phi}^2 = 	0 \,,
	\end{equation}
where the torsion scalar $T$ is at least $O(4)$, 
and $V''\equiv d^2\phi/d\phi^2$ and so forth. 
The lowest order of Eq.~(\ref{eq: scalar field eq}) indicates $V'(\phi_0)=0$, while at $O(2)$ we have
	\begin{equation}
	\label{eq: scalar order 2}
		\nabla^2\delta\phi^{(2)} + V''(\phi_0)\delta\phi^{(2)}=0.
	\end{equation}
Note that the perturbation caused by the matter in the Solar System should vanish 
at the cosmological distance since we require that $\phi$ has its cosmological value 
at a distance far away from the source, $i.e.$ $\delta\phi^{(2)} \to 0$ for $|x-x'| \to 0$. Due to the absence of the
gravitational field and the matter source, one finds that
 Eq.~(\ref{eq: scalar order 2}) has the solution $\delta\phi^{(2)} = 0$.
 This is similar to the case of the Gauss-Bonnet theory \cite{GB} whose  scalar field has a nonminimal coupling with the topologically invariant gravity.    
To  $O(4)$, Eq.~(\ref{eq: scalar field eq}) takes the form
	\begin{equation}
		\nabla^2\delta\phi^{(4)}-\xi T \phi_0 + V''(\phi_0) \delta\phi^{(4)} = 0, 
	\end{equation}
which has the solution
	\begin{equation}
		 \delta\phi^{(4)} = {1\over 4\pi} \int {\xi T \phi_0 \over \lvert \mathbf{x-x'} \rvert}
		  e^{-\sqrt{V''(\phi_0)}  \lvert \mathbf{x-x' }\rvert }d^3x'.
	\end{equation}
	
On the other hand, we now rewrite the equation motion in
Eq.~(\ref{eq: nonminimal fieldeq}) in the same manner as:
\begin{equation}
\begin{aligned}
\label{eq: TDEfieldeq2}
e^a{}_\mu e^{-1}\partial_\rho(eS_a{}^{\rho\nu})-\frac{1}{2}\delta_\mu^\nu e^a{}_\sigma e^{-1}
\partial_\rho \left( eS_a{}^{\rho\sigma}  \right) &+S_\rho{}^{\nu\sigma}
T^\rho{}_{\sigma\mu}+\frac{1}{4}\delta_\mu^\nu S_\rho{}^{\sigma\lambda}T^\rho{}_{\sigma\lambda} \\
=& {4\pi G_{\text{eff}}\over c^4} \left(  \mathcal{T}_\mu{}^\nu-\frac{1}{2}\delta_\mu{}^\nu \mathcal{T} +  \mathcal{T}(\phi)_\mu{}^\nu -\frac{1}{2}\delta_\mu{}^\nu \mathcal{T}(\phi)\right),
\end{aligned}
\end{equation}
where we have defined 
$G_\text{eff} = G(1+ {8\pi G\over c^4}\xi\phi_0^2)^{-1} $, and
\begin{equation}
\mathcal{T}(\phi)_\mu{}^\nu=
\delta_\mu{}^\nu \bigg[ {1\over2}  \partial_\rho \phi  \partial^\rho \phi
		 - V(\phi)  \bigg] - \partial^\nu\phi\partial_\mu\phi
	+4\xi S_\mu{}^{\lambda \nu} \phi  \partial_\lambda\phi.  
\end{equation}
Remarkably, the vanishing value of $\delta\phi^{(2)}$ implies that 
the term $S_0{}^{\rho 0} \left(  \partial_\sigma \phi  \right)$
is led by $O(6)$. Since $S_i{}^{0 j}\sim O(3)$, we find that
the constraint in Eq.~(\ref{eq: TDE constaints})
never plays a role in the PPN limit of the teleparallel dark energy model. As a result, no extra degree of freedom occurs other than $\phi$, 
so that we may use Eq.~(\ref{eq: tetradperturbation}) to solve the
perturbation of the vierbein field. 

The $(i,j)$  and  $(0,i)$ components up to
$O(2)$ and $O(3)$ in the perturbation of Eq.~(\ref{eq: TDEfieldeq2}) are 
\begin{equation}
	\begin{aligned}
	\label{eq: ij fieldeq of DE}
	&\partial_\rho S_{i}{}^{\rho j}-{1\over2}\delta_i^j\partial_\rho S_\sigma{}^{\rho\sigma} 
		= \frac{4\pi G_{\text{eff}}}{c^4}  \left( \mathcal{T}_i{}^j-{1\over2} \delta_i{}^j\mathcal{T}\right),\\
	&\partial_\rho S_{0}{}^{\rho i}= 
	\frac{4\pi G_{\text{eff}}}{c^4}\,  \mathcal{T}_0{}^i ,
	\end{aligned}
\end{equation}
respectively,
while the contribution of $\mathcal{T}(\phi)_\mu{}^\nu$ only appears
in the $(0,0)$ component up to $O(4)$, which is
	\begin{equation}
	\begin{aligned}
	\label{eq: 00 fieldeq of DE}
	 e^a{}_0  e^{-1}\partial_\rho\left( e S_a{}^{\rho0} \right)  
	-{1\over2}e^a{}_\lambda e^{-1}\partial_\rho
	\left( e S_a{}^{\rho\lambda} \right) 
	+ S_\rho{}^{0\sigma} T^\rho{}_{\sigma0}&
	+{1\over 4} T \\
	= {4\pi G_{\text{eff}} \over c^4}
	\bigg[ \mathcal{T}_0{}^0-{1\over2}\mathcal{T} + {1\over2}\dot{\phi}_0^2 -V(\phi_0) \bigg].
	\end{aligned}
	\end{equation}
The resulting vierbein field of these equations is
\begin{equation}
	\begin{aligned}
{}^{(2)}B_{ij}&=-\frac{\tilde{U}}{c^2}\delta_{ij},\\
{}^{(3)}B_{0i}&={}^{(3)}B_{i0}
=\frac{1}{c^3}\left(-\frac{7}{4}\tilde{V}_i
-\frac{1}{4}\tilde{W}_i\right),\\
{}^{(2+4)}B_{00}&=-{\tilde{U}\over c^2} - {\tilde{U}^2\over2 c^4} - \frac{1}{c^4}
\left(2\tilde{\Phi}_1- 2\tilde{\Phi}_2 + \tilde{\Phi}_3 + 3\tilde{\Phi}_4 \right) 
	-\frac{2G_\text{eff}}{c^4}\int \frac{ \left( {1\over2}\dot{\phi}_0^2 
	- V(\phi_0)  \right) }{\lvert \mathbf{x-x'} \rvert}d^3x' ,
\end{aligned}
	\end{equation}
where the tilde means that the Newtonian constant in the potentials of
$U$, $\Phi$, $V_i$, and $W_i$ is replaced by $G_\text{eff}$.
Constraints from cosmological observations to the non-minimally coupled
theory in Eq.~(\ref{eq: nonminimal action}) for various potentials 
(including $V(\phi)=0$) indicate the range of the best-fit
value $\lvert\xi\rvert\sim 10^{-1}-10^0$ \cite{Geng:2011aj}. 
Therefore, we have the typical value of the difference
$\lvert G/G_\text{eff}-1\rvert<10^{-10}$.
Defining that $r=\lvert \mathbf{x-x'} \rvert$, the corresponding metric perturbation reads
\begin{equation}
	\begin{aligned}
	\label{eq: h00 TDE}
{}^{(2)}h_{ij}&=-\frac{2\tilde{U}}{c^2}\delta_{ij},\\
{}^{(3)}h_{0i}&={}^{(3)}h_{i0}
=\frac{1}{c^3}\left(-\frac{7}{2}\tilde{V}_i
-\frac{1}{2}\tilde{W}_i\right),\\
{}^{(2+4)}h_{00}&=-{2\tilde{U}\over c^2} - {2\tilde{U}^2\over2 c^4} -\frac{1}{c^4}
\left(4\tilde{\Phi}_1- 4\tilde{\Phi}_2 + 2\tilde{\Phi}_3 + 6\tilde{\Phi}_4 \right)
	- \frac{8\pi G_\text{eff}  }{c^4}\left( {1\over2}\dot{\phi}_0^2 
	- V(\phi_0)  \right)r^2 .
\end{aligned}
	\end{equation}
Note that ${1\over2}\dot{\phi}_0^2 - V(\phi_0)$ has a energy density of
the same order as the cosmological constant $\sim H_0^2/\kappa^2$, 
and it has no contribution to the Solar System observations.
By neglecting the term with ${1\over2}\dot{\phi}_0^2 - V(\phi_0)$
in Eq.~(\ref{eq: h00 TDE}),
one recovers the standard results in the minimal-coupling limit
$\xi=0$ where $G_\text{eff}=G$.

Given that one may rescale the background value to $\phi_0=0$ 
so that $G_{\text{eff}}=G$, the effective Newtonian constant
shows no effect on the observational parameters. Hence, we find
that Eq.~(\ref{eq: h00 TDE}) gives $\gamma=\beta=1$ and zero
for the rest PPN parameters. This indicates that the teleparallel
dark energy model is indistinguishable from TEGR in the Solar System
scale up to the post-Newtonian order.


\section{Conclusions}
We have studied the weak-field regime in teleparallel gravity from a Minkowski background expansion.
Without  loss of generality, we have chosen a frame in which  the Sun is at rest in both  coordinate and  orthonormal bases.
In the Newtonian limit, the equation of motion for a slow-moving object is governed by the torsion
generated by the source matter with the perturbed vierbein field $B_{00} \simeq -U/c^2$.
To proceed the post-Newtonian calculation, we have chosen a vierbein expansion with the diagonal
Newtonian limit, which determines the components  $B_{ij}$,  $B_{0i}$
and $B_{00}$ up to $O(2)$, $O(3)$ and  $O(4)$, respectively.
By applying this formalism into TEGR, we have reproduced the standard PPN metric perturbation
as that in general relativity. 

We have also considered the post-Newtonian limit of the teleparallel dark energy model, $i.e.$
 TEGR with a scalar field which non-minimally couples to gravity. 
 Since the  torsion scalar
 $T\sim O(4)$ leads to the vanish of the perturbation of the scalar field ${}^{(2)}\delta\phi=0$,
the constraint equation in Eq.~(\ref{eq: TDE constaints}) has no contribution to the PPN
study, and the cosmic scalar $\phi$ is the only new degree of freedom in the teleparallel dark energy model.
Apart from the correction related to ${1\over2}\dot{\phi}_0^2 - V(\phi_0)$ at the
same order as the energy density of the cosmological constant, 
 we have shown that  $\beta=\gamma=1$ and zero for the rest of the other PPN parameters.
These results are independent of the mass or the potential energy of the scalar field.

It should be emphasized that our results in this work are different from 
those in 
scalar-tensor theories~ \cite{Olmo}.
In general,
when the non-minimally coupled scalar field receives a mass $m$ from its potential, the Newtonian potential
is modified to a Yukawa type $U(r)=Ue^{-mr}$, and the PPN parameter shows a dependence of the
distance from the source matter $\gamma=\gamma(r)$. The reason is that the source matter involves in the
solution of the $O(2)$ perturbation of the scalar field.

Taking the trace of Eq.~(\ref{eq: TEGR fieldeq}), we have found that
$\kappa^2\mathcal{T}=e^{-1}\partial_\mu(eS_\nu{}^{\mu\nu})$, while the source matter only responds to
the derivative of the torsion tensor. Given that the torsion scalar $T$ is quadratic of the torsion tensor, we can see from
Eq.~(\ref{eq: scalar field eq}) that the matter source only contributes to the solution of the perturbed scalar field
in an indirect way. Therefore, 
through a rescaling of the cosmological background, the Newtonian potential does not get 
affected by the arbitrary potential of the non-minimally coupled scalar field and 
the dark energy model in teleparallel gravity satisfies 
with Solar System constraints. 

\begin{acknowledgments}
We would like to thank Keisuke Izumi and Yen-Chin Ong for useful discussions.
The work was supported in part by National Center for Theoretical Sciences,  National Science
Council (NSC-98-2112-M-007-008-MY3 and NSC-101-2112-M-007-006-MY3) and
National Tsing-Hua University (102N2725E1),  Taiwan, R.O.C.
\end{acknowledgments}

\section*{Appendix A}
To compute the trace of the energy momentum tensor 
$\mathcal{T}=g^{\mu\nu}\mathcal{T}_{\mu\nu}$, we need  to
obtain the inverse metric $g^{\mu\nu}=\eta^{ab}e_a{}^\mu e_b{}^\nu$.
We shall define
\begin{equation}
		e_a{}^\mu=\delta_a{}^\mu + {}^{(2)}C_a{}^\mu +{}^{(3)}C_a{}^\mu+{}^{(4)}C_a{}^\mu+ \ldots\,,
	\end{equation}
which leads to the decomposition $g^{\mu\nu}=\eta^{\mu\nu}+h^{\mu\nu}$.
Using the condition in Eq.~(\ref{eq: orthogonality}), we find that
to $O(4)$:
\begin{eqnarray}
    \label{eq: inverse vierbein}
{}^{(2+3)}C_a{}^\mu&=&-\delta_a{}^\lambda{}^{(2+3)}B^\mu{}_\lambda,\\
{}^{(4)}C_a{}^\mu&=&-\delta_a{}^\lambda{}^{(4)}B^\mu{}_\lambda
+\delta_a{}^\lambda{}^{(2)}B^\mu{}_\nu{}^{(2)}B^\nu{}_\lambda.   
    \end{eqnarray}
Given that the source matter vanishes at the background level, 
the relevant trace is given by $\mathcal{T}=(\eta^{\mu\nu}+{}^{(2)}h^{\mu\nu})\mathcal{T}_{\mu\nu}$,
where
\begin{equation}
{}^{(2)}h^{\mu\nu}=-\eta^{\mu\lambda}\,{}^{(2)}B^\nu{}_\lambda-\eta^{\lambda\nu}\,{}^{(2)}B^\mu{}_\lambda.
\end{equation} 

\section*{Appendix B}

We calculate to the solution of $B^0{}_0$ up to $O(4)$ by
setting $\mu=\nu=0$ to Eq.~(\ref{eq: fieldeq2}), which becomes
	\begin{equation}
	\label{eq: post00}
		\underbrace{ e^a{}_0 e^{-1}\partial_\sigma \left(e S_a{}^{\sigma 0}\right) }_\text{Part A} 
		-{1\over2}\underbrace{ \delta^0_0 e^a{}_\rho e^{-1}\partial_\sigma\left(e S_a{}^{\sigma\rho}\right)}		_\text{Part B} 
		+\underbrace{S_\rho{}^{0\sigma} \torsion{\rho}{\sigma}{0}}_\text{Part C}
		+{1\over4} \underbrace{  \delta_0^0 \s{\rho}{\sigma}{\lambda} T_{\rho}{}^{\sigma\lambda}}_\text		{Part D}
		= {4\pi G\over c^4} \left( T_0{}^{0} -{1\over2} \delta^0_0 T  \right),
	\end{equation}
where Parts A and B contain linear terms of vierbein at $O(2)$ 
and nonlinear ones of at least $O(4)$, 
while Parts C and D are no lower than $O(4)$ 
since they are composed of the square of the torsion tensor. 
We  compute the left-hand side of Eq.~(\ref{eq: post00}) in terms of the Newtonian gravitational potential $U$.
For  Part A, we have
	\begin{eqnarray}
		e^a{}_0 e^{-1}\partial_\sigma \left(e S_a{}^{\sigma 0}\right) 
		&
		= &\partial_\sigma S_0{}^{\sigma 0} - S_a{}^{\sigma 0} \partial_\sigma \B{a}{0} 
		+  S_0{}^{\sigma 0} \partial_\sigma e \nonumber\\
		&=&\partial_\sigma S_0{}^{\sigma 0}- \left( U\nabla^2U - {1\over2}\nabla^2U^2  \right) 
		+ \left			( \nabla^2U^2-2U\nabla^2U \right)\nonumber\\
		&=&\partial_\sigma  S_0{}^{\sigma 0} - 3U\nabla^2U + {3\over2}\nabla^2U^2,
	\end{eqnarray}
and for Part B
	\begin{eqnarray}
		\delta^0_0 e^a{}_\rho e^{-1}\partial_\sigma\left(e S_a{}^{\sigma\rho}\right) 
		&=& e^{a}{}_\rho\partial_		\sigma S_a{}^{\sigma\rho} 
		+ \delta^a_\rho S_a{}^{\sigma\rho}\partial_\sigma e \nonumber\\
		&=& \left( \partial_\sigma S_\rho{}^{\sigma\rho} - U\nabla^2U
		+{1\over2}\nabla^2U^2 \right) + \left		( \nabla^2U^2-2U\nabla^2U \right)\nonumber\\
		&=& \partial_\sigma S_\rho{}^{\sigma\rho} +{3\over2}\nabla^2U^2-3U\nabla^2U.
	\end{eqnarray}
Combining these two terms, one obtains a more simpler form
	\begin{eqnarray}
		&& e^a{}_0 e^{-1}\partial_\sigma \left(e S_a{}^{\sigma 0}\right)
		-{1\over2} \delta^0_0 e^a{}_\rho e^{-1}\partial_\sigma\left(e S_a{}^{\sigma\rho}\right) \nonumber\\
		&=& \left( \partial_\sigma S_0{}^{\sigma 0} -{1\over2}\partial_\sigma S_\rho{}^{\sigma\rho}  \right)
				+{3\over4}\nabla^2U^2 - {3\over2}U\nabla^2U\nonumber\\
		&=&  \left( -{1\over2}\nabla^2 \B{0}{0} - {1\over4}\nabla^2U^2 \right)
		+{3\over4}\nabla^2U^2 - 		{3\over2}U\nabla^2U.
	\end{eqnarray}
Similarly, Parts  C and D are given by
	\begin{equation}
		 S_\rho{}^{0\sigma} \torsion{\rho}{\sigma}{0}= U\nabla^2U-{1\over2}\nabla^2U^2,
	\end{equation}
and
	\begin{equation}
		\delta_0^0 \s{\rho}{\sigma}{\lambda} T_{\rho}{}^{\sigma\lambda}= \nabla^2U^2-2U\nabla^2U.
	\end{equation}
Together with the matter source Eq.~(\ref{eq: em expansion}), 
the $(0, 0)$ component of the field equation becomes
	\begin{equation}
		-{1\over2}\nabla^2 \B{0}{0} + {1\over4}\nabla^2 U^2 - U\nabla^2 U 
		= -{2\pi G\over c^4} \left( \rho c^2+ \rho\Pi + 2\rho v^2 +3p \right).
	\end{equation}
Utilizing the potentials of Eqs.~(\ref{potential Phi}),
the vierbein expanded to the post-Newtonian order is
	\begin{equation}
		\B{0}{0} = {U\over c^2} + {U^2\over2 c^4} + \frac{1}{c^4}\left(2\Phi_1- 2\Phi_2 + \Phi_3 + 3\Phi_4 \right). 
	\end{equation}


\end{document}